\documentclass[twocolumn, preprint]{aastex631}

\usepackage{xcolor}
\usepackage{amsmath,amssymb}
\usepackage{orcidlink}
\usepackage{hyperref}

\begin{document}

\title{The Non Parametric Reconstruction of Binary Black Hole Mass Evolution from GWTC-4.0 Gravitational Wave Catalog}
\author{Samsuzzaman Afroz \orcidlink{0009-0004-4459-2981}}
\email{samsuzzaman.afroz@tifr.res.in}
\author{Suvodip Mukherjee \orcidlink{0000-0002-3373-5236}}
\email{suvodip@tifr.res.in}
\affiliation{Department of Astronomy and Astrophysics, Tata Institute of Fundamental Research, Mumbai 400005, India}

\begin{abstract}
The distribution of binary black hole (BBH) masses and its evolution with redshift provide key insights into the different formation channels of compact objects and their dependence on cosmic time and stellar properties such as metallicity and star formation history. We present a non-parametric framework of the redshift evolution of the BBH mass distribution using the gravitational-wave (GW) catalogs GWTC-3 and GWTC-4 of LIGO-Virgo-KAGRA (LVK). This method simultaneously searches for the signature of any linear and quadratic redshift evolution with respect to the low-redshift population in a Bayesian framework, taking into account the detector selection effects. We find tentative evidence for a linear redshift-dependent evolution of the mass distribution, consistent over a mass range ($m \gtrsim 50\,M_\odot$), while lower-mass systems show no significant evolution. The quadratic term remains consistent with zero, indicating that a simple linear dependence adequately describes the population up to redshift $z \sim 1$. With more GW sources in the future, this technique can reveal subtle evolutionary features in BBH populations and provide new insights into the cosmic history of black hole formation.
\end{abstract}

\section{Introduction}

The detection of gravitational waves (GWs) has revolutionized astrophysics by providing a direct means to observe and characterize binary black hole populations across vast cosmic distances. These unprecedented observations have transformed our understanding of how stellar mass black holes form, evolve, and merge over the history of the universe \citep{LIGOScientific:2018mvr,LIGOScientific:2016aoc,Bouffanais:2021wcr,Franciolini:2021tla,Cheng:2023ddt,Antonelli:2023gpu,2023ApJ...950..181W,Tiwari:2020otp,Tiwari:2021yvr, Tiwari:2025lit, Kimball:2020opk, Kimball:2020qyd,Afroz:2024fzp, Afroz:2025efn, Afroz:2025ikg}. With each new detection, we gain invaluable information about the masses, spins, and rates of these elusive objects, enabling tests of astrophysical theory in regimes previously inaccessible \citep{LIGOScientific:2020kqk, KAGRA:2021duu, LIGOScientific:2025pvj}.

A fundamental question arising from these advances is whether the underlying distribution of binary black hole (BBH) masses, called as the BBH mass function changes over cosmic time. Investigating this evolution is crucial because it encodes rich information about the life cycles of massive stars, the influence of their environments, and the physical processes shaping their end states \citep{Mukherjee:2021rtw,Bailes:2021tot, Arimoto:2021cwc,Mapelli:2021taw,Karathanasis:2022rtr,Iorio:2022sgz,Rinaldi:2025emt,Barrett:2017fcw,Dominik:2012kk,Bailyn:1997xt,Ye:2024ypm,Schiebelbein-Zwack:2024roj,Rinaldi:2023bbd,Gennari:2025nho,Lalleman:2025xcs,Sicilia:2021gtu}.

Factors such as the chemical composition of progenitor stars, commonly referred to as metallicity, vary significantly across epochs and profoundly impact stellar winds and mass loss rates. Lower metallicities prevalent in the early universe likely allowed stars to retain more mass, potentially leading to the formation of heavier black holes \citep{Karathanasis:2022rtr,Farrell:2020zju,Belczynski:2019fed,Spera:2017fyx,Farmer:2020xne,Hendriks:2023yrw,Farmer:2019jed,Leung:2019fgj,Tong:2025wpz, Antonini:2025ilj,Costa:2022aka, Fishbach:2017zga, Talbot:2018cva}. Moreover, the universe’s first generation of stars, the Population III stars, constituted metal free environments that could have produced exceptionally massive black holes distinct from those observed today \citep{Santoliquido:2023wzn,Costa:2023xsz,Giacobbo:2017qhh,Dvorkin:2016wac}.

Understanding how these astrophysical processes imprint themselves on the black hole mass function as a function of redshift holds the promise of uncovering the origins and evolution of compact object populations, offering insights into star formation history, galaxy evolution, and the end stages of stellar lives. Previous studies have explored various approaches to characterize the BBH mass distribution and its potential evolution, ranging from parametric models with specific functional forms to more flexible non-parametric methods \citep{Wysocki:2018mpo,Mandel:2018mve,Tiwari:2020otp,Talbot:2019okv,Wong:2020jdt,Callister:2023tgi,Riley:2023jep,Farah:2024xub,Talbot:2024yqw,Ray:2023upk,MaganaHernandez:2025fkm,Antonini:2025zzw,Edelman:2022ydv,Heinzel:2024hva,Heinzel:2023vkq,Heinzel:2024jlc,Alvarez-Lopez:2025ltt}.

The two major open questions regarding the redshift evolution of the BBH mass distribution are: (i) Does a single, global BBH mass distribution adequately describe the population across all redshifts, or is such a description valid only within restricted redshift ranges? (ii) If not, does the BBH mass distribution itself evolve with redshift, and if so, what is the nature of that evolution? The challenge in answering these questions lies with our ignorance in modelling either of these from first principle due to large astrophysical uncertainties. Along with the theoretical uncertainties, selection effects also plays a major role.

In order to infer the true physical effect from the data, one needs to take into account the redshift and mass dependent selection effects. Gravitational wave detectors have sensitivity biases favoring the detection of more massive and nearby sources, creating a mass and redshift dependent selection function effect, which can be modeled given the detectors sensitivity \citep{PhysRevD.108.043011, PhysRevD.110.103018, Messenger:2012jy,Gerosa:2024isl,Essick:2025zed}.

In this work, we propose a fully non parametric framework for inferring the evolution of the stellar mass black hole mass function using GW observations. Leveraging a Taylor series expansion of the joint mass redshift distribution, this approach makes minimal assumptions about the functional form of the mass function or its evolution. Instead, it allows the data itself to reveal evolutionary patterns, capturing both linear and higher order changes in the population over redshift. This strategy offers greater flexibility and interpretability, with each Taylor coefficient corresponding to physically meaningful rates and accelerations of population change. Such a framework paves the way for revealing subtle and complex evolutionary signatures that could otherwise remain hidden. Applying this framework to the GWTC-3 \citep{KAGRA:2021duu} and GWTC-4 catalog \citep{LIGOScientific:2025slb}, the most comprehensive collection of BBH mergers to date, we find tentative evidence for mass dependent evolution in the BBH mass function. High mass systems ($40$-$60\,M_\odot$) very mild indications of positive redshift evolution, while lighter black holes exhibit little or no statistically significant evolution.

These results are consistent with metallicity driven formation channels and demonstrate the potential of non-parametric methods for probing black hole populations with future GW observations.

\section{Non-Parametric Bayesian Inference of the BBH Mass Distribution with Redshift}
\label{sec:Framework}

The aim of this work is to test whether the distribution of BBH component masses evolves with redshift, a question with direct implications for stellar evolution, metallicity, and the astrophysical environments of BBH formation. Gravitational wave observations provide unique access to this distribution, but detector sensitivity introduces strong selection effects: massive systems are detectable at greater distances than lighter ones, biasing the observed sample. Correcting for these biases is essential for reconstructing the intrinsic population. We demonstrate in this section our Bayesian framework for the reconstruction of the BBH mass distribution from data using a flexible, non-parametric approach.

\subsection{Non-Parametric Mass-Redshift Distribution}\label{sec:massredshift}

Rather than relying on parametric models with fixed functional forms, which risk obscuring unanticipated features, we adopt a non-parametric framework based on a Taylor expansion of the mass-redshift distribution. Instead of prescribing exact shapes, this method expresses $\rm{p(m,z)}$ as a smooth series expansion in redshift around a reference redshift $\rm{z_{\rm ref}}$:
\begin{equation}
    \rm{p(m,z) = \sum_{n=0}^N \frac{1}{n!} \left. \frac{\partial^n p(m,z)}{\partial z^n} \right|_{z=z_{\rm ref}} (z -z_{\rm ref})^n},
    \label{eq:taylorseries}
\end{equation}
where the truncation order $\rm{N}$ dictates the level of evolutionary behavior captured, and we choose $\rm{z_{ref} = 0.25}$ to center the expansion. In this work, we restrict our analysis to the second-order term in the Taylor series expansion, justified by the fact that the redshift range of current GW observations extends up to $\rm{z \sim 1.0}$, with only a few detected events at high redshift. The resulting approximation is:
\begin{equation}
    \rm{p(m,z) \approx p_0(m) + p_1(m) (z-z_{\rm ref}) + p_2(m) (z-z_{\rm ref})^2},
\end{equation}
with the coefficient functions defined as:
\begin{align}
\rm{p_0(m)\equiv p(m,z_{\rm ref})}, & \quad
\rm{p_1(m)\equiv\left.\frac{\partial p}{\partial z}\right|_{z_{\rm ref}}}, \nonumber \\ \quad
\rm{p_2(m)\equiv\frac{1}{2}\left.\frac{\partial^2 p}{\partial z^2}\right|_{z_{\rm ref}}}.
\label{eq:taylorcoeff}
\end{align}

Here, $\rm{p_0(m)}$ encodes the mass distribution at the reference redshift, while $\rm{p_1(m)}$ and $\rm{p_2(m)}$ capture the linear and quadratic evolutionary trends, respectively. This construction avoids restrictive parametric assumptions, permits non-monotonic evolution, and provides coefficients with direct physical interpretations. In practice, $\rm{p_0(m_k)}$, $\rm{p_1(m_k)}$, and  $\rm{p_2(m_k)}$ are evaluated at each mass bin $\rm{m_k}$ as described in Section~\ref{sec:hierinf}.

Since the Taylor expansion does not automatically satisfy probability normalization, in our analysis we explicitly normalize the resulting expression at each redshift by dividing by its integral over mass. This ensures that $\rm{\int p(m \mid z, \boldsymbol{\theta}) \, dm = 1}$ at each redshift, making $\rm{p(m \mid z, \boldsymbol{\theta})}$ a valid conditional probability distribution for mass given redshift. The recovered evolution coefficients encode distinct aspects of mass function evolution. The linear evolution coefficient $\rm{p_1(m)}$ indicates whether black holes of mass $\rm{m}$ were more common (positive values) or less common (negative values) at higher redshifts. The magnitude $\rm{|p_1(m)|}$ quantifies the strength of evolution. The quadratic evolution coefficient $\rm{p_2(m)}$ captures curvature in evolutionary trends. Positive values of $\rm{p_2(m)}$ indicate accelerating evolution, while negative values suggest deceleration. This term allows for non-monotonic behavior and can reveal whether evolutionary processes are intensifying or saturating with redshift.

Throughout this work, the primary mass $\rm{m_1}$ and the secondary mass $\rm{m_2}$ are the two labeled components of each binary, ordered by the convention $\rm{m_1 \geq m_2}$. They are therefore not statistically independent: the labeling itself correlates them, and there exists a single underlying joint distribution $\rm{p(m_1,m_2 \mid z)}$. In this work we do not reconstruct this joint distribution. Instead, we reconstruct its two one-dimensional marginals,

\begin{equation}
\begin{split}
\rm{p(m_1 \mid z) = \int p(m_1,m_2 \mid z)\, dm_2}, \\
\rm{p(m_2 \mid z) = \int p(m_1,m_2 \mid z)\, dm_1},
\end{split}
\end{equation}

by applying the framework of Eqs.~(\ref{eq:taylorseries})--(\ref{eq:taylorcoeff}) \emph{separately} to each component. For the $\rm{m_1}$ analysis, $\rm{m}$ denotes the primary mass and the primary-mass posterior samples of each event are used; for the $\rm{m_2}$ analysis, $\rm{m}$ denotes the secondary mass and the secondary-mass posterior samples are used. Eq.~(\ref{eq:taylorseries}) is thus a one-dimensional expansion applied to each marginal in turn, not a multi-dimensional expansion over $\rm{(m_1,m_2)}$ jointly. We emphasize that ``separately'' here refers to running two distinct hierarchical analyses and does \emph{not} imply an assumption of statistical independence between $\rm{m_1}$ and $\rm{m_2}$.

Crucially, each marginal distribution of masses is used with its own marginal selection function of the corresponding masses. The primary-mass analysis uses $\rm{S(m_1,z)}$, obtained by binning recovered injections in $\rm{(m_1,z)}$ with the secondary mass integrated out over the injection population; the secondary-mass analysis uses the distinct function $\rm{S(m_2,z)}$, obtained by binning recovered injections in $\rm{(m_2,z)}$ with the primary mass integrated out. Reusing a single component-mass selection function for both would be inconsistent, since the detectability of a binary at fixed $\rm{m_2}$ depends on the marginalized distribution of its partner $\rm{m_1}$, and vice versa.

Our goal is to reconstruct, in a purely data driven way, how the distribution of each component mass evolves with redshift. This separate treatment of the two marginals is valid precisely because we make no assumption about the formation channel or the physical mechanism relating the two component masses. Since we do not model how the two masses are jointly produced, we consider the distribution of the individual component mass, marginalized over the other. So, that the mass distribution for each component can be inferred as a function of redshift. When analyzing one component mass, the partner mass is marginalized over its LVK PE prior (Section~\ref{sec:hierinf}), and the inferred evolution of each marginal is conditional on this choice. Any similarity or difference in how the two marginals evolve therefore arises from the data, given this choice, rather than from an imposed relation between the two component masses.

\subsection{Data Processing and Selection Function}

We analyze BBH mergers from the GWTC-3 \citep{KAGRA:2021vkt} and GWTC-4 \citep{LIGOScientific:2025slb} catalogs, using publicly released posterior samples from the LVK parameter estimation analyses. For each detected event $i$ and posterior sample $s$, we obtain source-frame component masses $(m_{1,s}^{(i)}, m_{2,s}^{(i)})$ and a luminosity distance sample $d_{L,s}^{(i)}$. Adopting a spatially flat $\Lambda$CDM cosmology with parameters from \textsc{Planck} 2018 \citep{Planck:2018vyg}, the luminosity distance as a function of redshift is:
\begin{equation}
d_L(z) = (1+z)\,\frac{c}{H_0}\int_{0}^{z}\frac{dz'}{\sqrt{\Omega_m(1+z')^3 + (1-\Omega_m)}}, 
\end{equation}
where $\Omega_m$ is the matter density and $H_0$ is the Hubble constant. We convert each distance sample $d_{L,s}^{(i)}$ to a redshift $z_s^{(i)}$ using this relation, mapping the posteriors into the $(m,z)$ plane. We work consistently in the source frame for easier interpretation of any redshift evolution signatures.

To compute the selection function, we use the publicly released LVK injection campaigns associated with GWTC-3 and GWTC-4 

\footnote{The injection sets and recovered selection functions for GWTC-3 and GWTC-4 are publicly available on Zenodo: \citet{GWTC3_injections} and \citet{GWTC4_injections}, respectively.}

These injection campaigns provide simulated compact binary coalescence signals spanning the relevant parameter space. We classify an injection as detected if its network signal-to-noise ratio satisfies $\mathrm{SNR} > 10$.

We note that this represents a mass-averaged approximation that roughly corresponds to a $\sim\!1$\,yr$^{-1}$ false-alarm rate \citep{Essick:2023toz}, rather than the FAR-based detection criterion actually used in GW searches. In practice, the effective SNR threshold varies as a function of mass, and neglecting these mass-dependent variations could in principle introduce small systematic biases in our inferred mass-dependent evolution coefficients. We have explicitly tested for this effect by repeating the full analysis with a FAR-based selection function ($\rm{FAR < 1\,yr^{-1}}$); as shown in Appendix~\ref{app:far_vs_snr}, the resulting evolution coefficients are statistically consistent with those from the SNR-based analysis, leaving all of our conclusions unchanged.

The selection function $\rm{S(m,z)}$ represents the probability that a binary with component mass $\rm{m}$ at redshift $\rm{z}$ is detected by the GW network. Formally, this corresponds to marginalizing the detection probability over extrinsic parameters $\rm{\Omega}$ (e.g., sky location, inclination, and polarization), as well as the intrinsic parameters spanned by the injection campaign, including spin magnitudes, spin orientations, and mass ratios:
\begin{equation}
\rm{S(m,z) = \int P(\mathrm{det} \mid m, z, \Omega) \, p(\Omega) 
\, d\Omega \;\approx\; \frac{N_{\mathrm{det}}(m,z)}
{N_{\mathrm{inj}}(m,z)}},
\end{equation}
where the integral is estimated empirically using the injection campaigns described above. We stress that the generic notation $\rm{S(m,z)}$ used here stands for either the primary-mass selection function $\rm{S(m_1,z)}$ or the secondary-mass selection function $\rm{S(m_2,z)}$, depending on which marginal distribution is being analyzed. The former is obtained by binning recovered injections in $\rm{(m_1,z)}$ with the secondary mass integrated out over the injection population, and the latter by binning in $\rm{(m_2,z)}$ with the primary mass integrated out; the two are distinct functions and are never interchanged. This is consistent with the separate treatment of the two component masses described at the end of Section~\ref{sec:massredshift}. In practice, $\rm{S(m,z)}$ is computed as the fraction of injected signals that are recovered in each region of the $\rm{(m,z)}$ plane.

We note that the injection distributions over spins are designed for broad parameter coverage rather than to follow a specific physical population. This implies that $S(m,z)$ implicitly adopts the injected spin distribution, which may introduce a small systematic uncertainty in the inferred selection function. We verify explicitly in Appendix~\ref{app:spin_reweight} that reweighting the injected spins to a plausible astrophysical population leaves our inferred evolution coefficients unchanged.

To avoid numerical instabilities in sparsely populated regions of the injection  parameter space, we impose a small floor on the selection function and require $\rm{S(m,z) \geq 10^{-3}}$. Because the injection sets span realistic distributions of mass ratios, spins, and orientations, the resulting selection function automatically incorporates spin- and geometry-dependent detection efficiencies. By working directly with the component mass posteriors $\rm{(m_1, m_2)}$, we effectively marginalize over these secondary parameters without imposing restrictive parametric assumptions about their redshift evolution.

\subsection{Hierarchical Inference Framework}
\label{sec:hierinf}

We develop a hierarchical Bayesian framework that properly accounts for measurement uncertainties in individual event parameters and detector selection effects. Our approach builds on the standard hierarchical inference methodology established in gravitational wave population studies \citep{Mandel:2018mve,Fishbach:2017dwv,Loredo:2004nn}, adapted to our non-parametric, binned representation of the mass-redshift distribution. We discretize the continuous $\rm{(m,z)}$ parameter space into a two-dimensional grid of bins $\rm{\{(m_k, z_j)\}}$ with $\rm{k = 1, \ldots, N_m}$ mass bins and $\rm{j = 1, \ldots, N_z}$ redshift bins.

This bin approach minimizes assumptions about smoothness or correlations between neighboring regions, allowing the data to reveal localized evolutionary features without imposing artificial constraints. The intrinsic number of mergers in bin $\rm{(m_k, z_j)}$ is:

\begin{align}
\rm{N_{\rm int}(m_k, z_j \mid \boldsymbol{\theta}_{kj})} & = \rm{T_{\rm obs} \, \mathcal{R}(z_j) \, p(m_k \mid z_j, \boldsymbol{\theta}_{kj})} \, \nonumber \\ &\times \rm{\frac{dV_c}{dz}\bigg|_{z_j} \, \frac{1}{1+z_j} \, \Delta m_k \, \Delta z_j},
\label{eq:Nint}
\end{align}

where $\rm{T_{\rm obs}}$ is the observation time, $\rm{\mathcal{R}(z_j)}$ is the merger rate density, $\rm{dV_c/dz}$ is the differential comoving volume, and $\rm{(1+z)^{-1}}$ accounts for time dilation. This represents mergers that \emph{actually occur} in the universe, independent of detection. The expected number of detections is $\rm{N_{\rm exp}(m_k, z_j \mid \boldsymbol{\theta}_{kj}) = N_{\rm int}(m_k, z_j \mid \boldsymbol{\theta}_{kj}) \times S(m_k, z_j)}$, where the selection function $\rm{S(m_k, z_j)}$ accounts for detector sensitivity. In practice, the merger rate density $\rm{\mathcal{R}(z_j)}$ is absorbed into the normalization of the mass distribution at each redshift bin, and our framework directly infers the conditional distribution $\rm{p(m_k \mid z_j)}$.

For each event $\rm{i}$ and bin $\rm{(k,j)}$, we compute the posterior weight:
\begin{equation}
\rm{W_{ijk} = \frac{1}{N_s} \sum_{\substack{\text{samples } \ell \\ \text{in bin } (k,j)}} \frac{1}{\pi(m_{i,\ell}, z_{i,\ell})}},
\label{eq:weights}
\end{equation}

where $\rm{\{m_{i,\ell}, z_{i,\ell}\}_{\ell=1}^{N_s}}$ are the posterior samples for event $\rm{i}$, and $\rm{\pi(m,z)}$ is the LVK parameter estimation (PE) prior on the analyzed mass and redshift. This weight represents the fractional posterior support of event $\rm{i}$ in bin $\rm{(k,j)}$. Since only the prior on the analyzed mass and redshift is divided out, the partner mass, spins, and extrinsic parameters of each event are marginalized over their LVK PE priors, without reweighting to any fiducial assumed population model.

The hierarchical likelihood for bin $\rm{(k,j)}$ with parameters $\rm{\boldsymbol{\theta}_{kj}}$ is:
\begin{equation}
\rm{\mathcal{L}(\boldsymbol{\theta}_{kj}) = e^{-N_{\rm exp}(m_k, z_j \mid \boldsymbol{\theta}_{kj})} \prod_{i=1}^{N_{\rm obs}} \left[ W_{ijk} \, \frac{N_{\rm int}(m_k, z_j \mid \boldsymbol{\theta}_{kj})}{\Delta m_k \, \Delta z_j} \right]}.
\label{eq:likelihood_main}
\end{equation}

here, the product is over the observed events. Here the division by $\Delta m_k \Delta z_j$ converts the intrinsic merger count $N_{\rm int}$ into a differential rate density, and the proportionality absorbs constant normalization factors independent of $\boldsymbol{\theta}_{kj}$.

\section{Observed BBH Population and Evidence for Mass Distance Trends}

\begin{figure*}[t]
    \centering
    \includegraphics[width=0.49\textwidth, height=6.5cm]{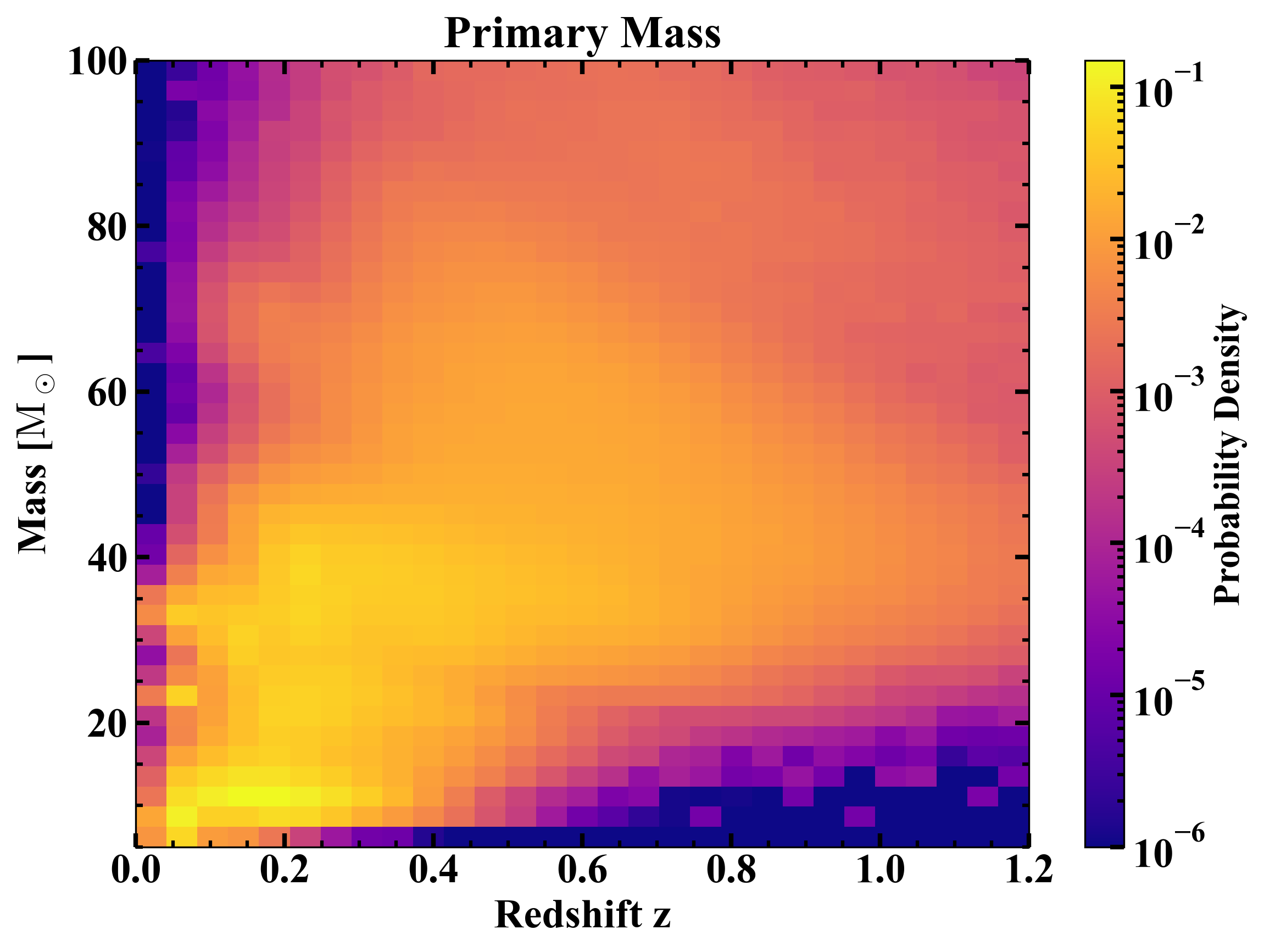}
    \includegraphics[width=0.49\textwidth, height=6.5cm]{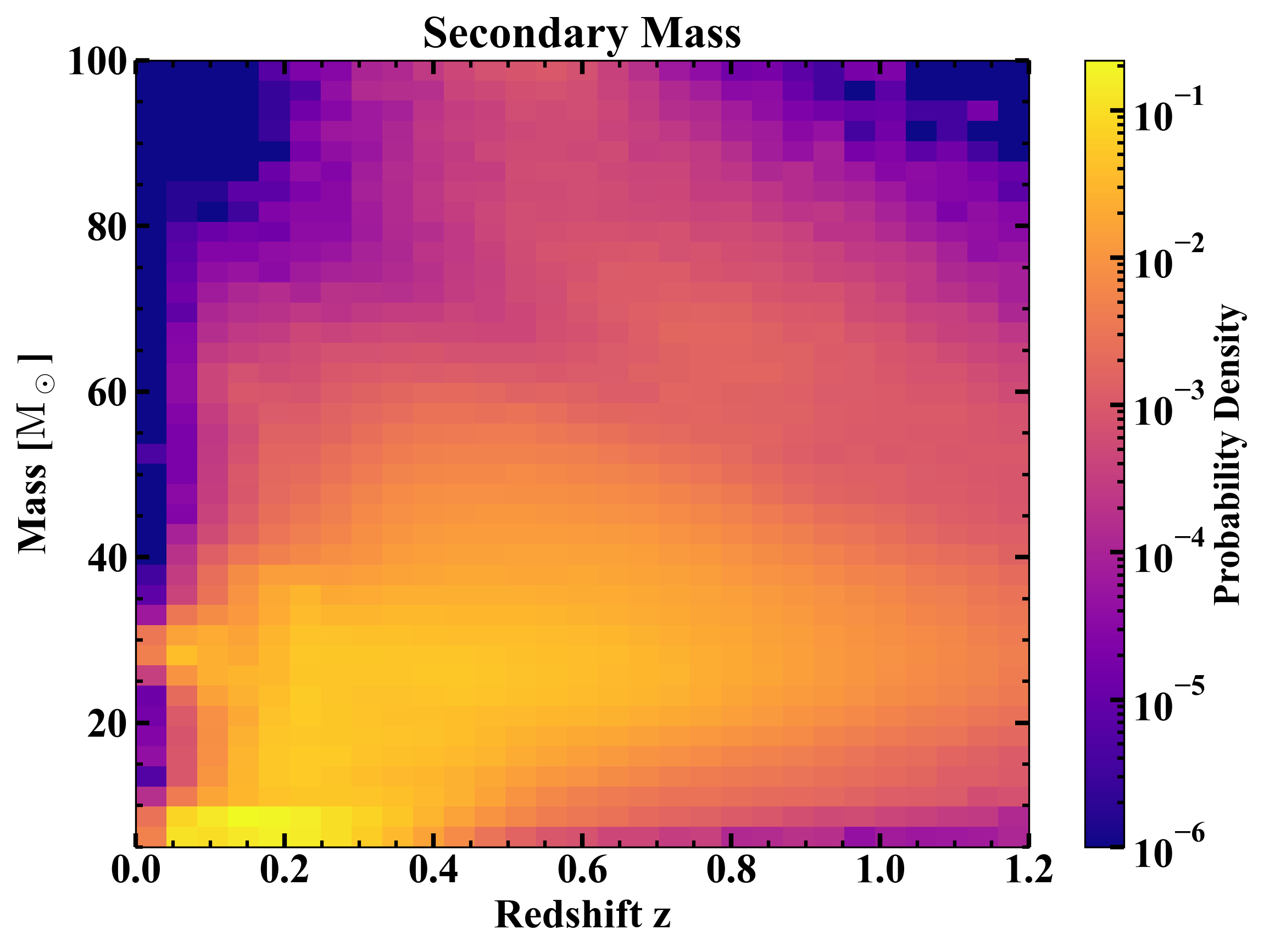}
    \caption{Mass-redshift distributions for the primary (left) and secondary (right) black hole components in GWTC-4 catalog. Each distribution is constructed from individual-event PE posterior samples, reweighted by the inverse of the PE prior $\pi(m,z)$ so that the resulting histogram reflects the likelihood contribution rather than the assumed PE prior.}
    \label{fig:raw_data}
\end{figure*}

We analyze BBH mergers from the GWTC-3 \citep{KAGRA:2021duu} and GWTC-4 \citep{LIGOScientific:2025slb} catalog, the most comprehensive public collection of GW detections to date. For each catalog event we draw posterior samples of the source frame component masses, $\rm{(m_1,m_2)}$, and luminosity distance, $\rm{d_L}$. Distances are converted to redshift assuming a spatially flat $\Lambda$CDM cosmology with \textsc{Planck}~2018 parameters ($\rm{H_0=67.4}\,$km\,s$^{ 1}$\,Mpc$^{ 1}$, $\rm{\Omega_m=0.315}$) \citep{Planck:2018vyg}, which maps every posterior sample into the $\rm{(m,z)}$ plane.

Figure~\ref{fig:raw_data} presents the resulting mass-redshift distributions of primary mass (left) and secondary mass (right) constructed from these posterior samples across all 176 events. Figure~\ref{fig:raw_data} presents the resulting mass-redshift distributions of primary mass (left) and secondary mass (right) constructed from these posterior samples across all 176 events. Each 2D histogram is constructed from PE posterior samples reweighted by the inverse of the PE prior $\pi(m,z)$, so that the resulting distribution reflects the data-driven likelihood contribution rather than the assumed PE prior, and can be used to visualize the intrinsic population trend. This reveals the characteristic BBH mass hierarchy and redshift evolution over $\rm{m\in[5,100]}\,$M$_\odot$, $\rm{z\in[0,1.2]}$.

The observed distribution of the GW sources in the mass and redshift plane can arise from two non exclusive effects. First, heavier binaries generate stronger GW signals and are therefore detectable at larger distances; this selection bias alone can produce an apparent mass redshift correlation even if the intrinsic population is non evolving. Second, there may be genuine astrophysical evolution of the BBH mass function for example, metallicity driven differences in stellar winds and mass loss, that enhances the formation of massive remnants at earlier cosmic times \citep{Fishbach:2021yvy}. Disentangling these two effects requires quantitative correction for detector sensitivity and measurement uncertainty.

\begin{figure*}[t]
    \centering    
    \includegraphics[width=1.00\textwidth, height=6.0cm]{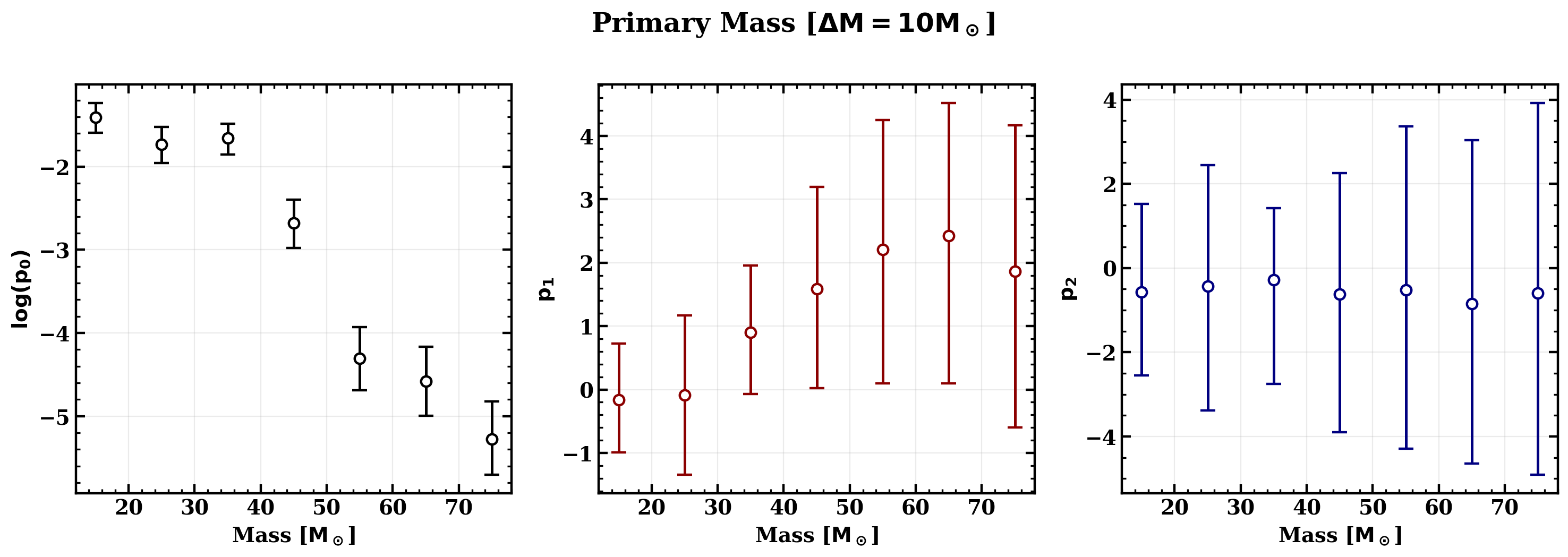}
    \caption{Posterior constraints on the redshift evolution of the BBH primary mass ($\rm{m_1}$) distribution for a binning scheme with $\rm{\Delta M = 10\,M_{\odot}}$ and $\rm{\Delta z = 0.1}$.  \textbf{Left:} Constant coefficient $\rm{\log(P_0)}$ describing the baseline primary mass distribution. \textbf{Middle:} Linear coefficient $\rm{P_1}$, quantifying the first-order variation of the mass distribution with redshift. \textbf{Right:} Quadratic coefficient $\rm{P_2}$, probing possible curvature in the redshift dependence.  Error bars represent $\rm{1\sigma}$ credible intervals obtained from the hierarchical Bayesian analysis.  The linear coefficient is consistent with zero at low masses ($\rm{m_1 \lesssim 30\,M_{\odot}}$) and shows a mild positive trend toward higher masses, suggesting that more massive black holes may be relatively more abundant at earlier cosmic times. The quadratic coefficient remains consistent with zero across all mass bins, indicating that a linear redshift dependence adequately describes the population within the current observational range $\rm{z \lesssim 1}$ accessible to LVK.}
    \label{fig:Mass1LargeBincoefficients}
\end{figure*}

\begin{figure*}[t]
    \centering
    \includegraphics[width=1.00\textwidth, height=6.0cm]{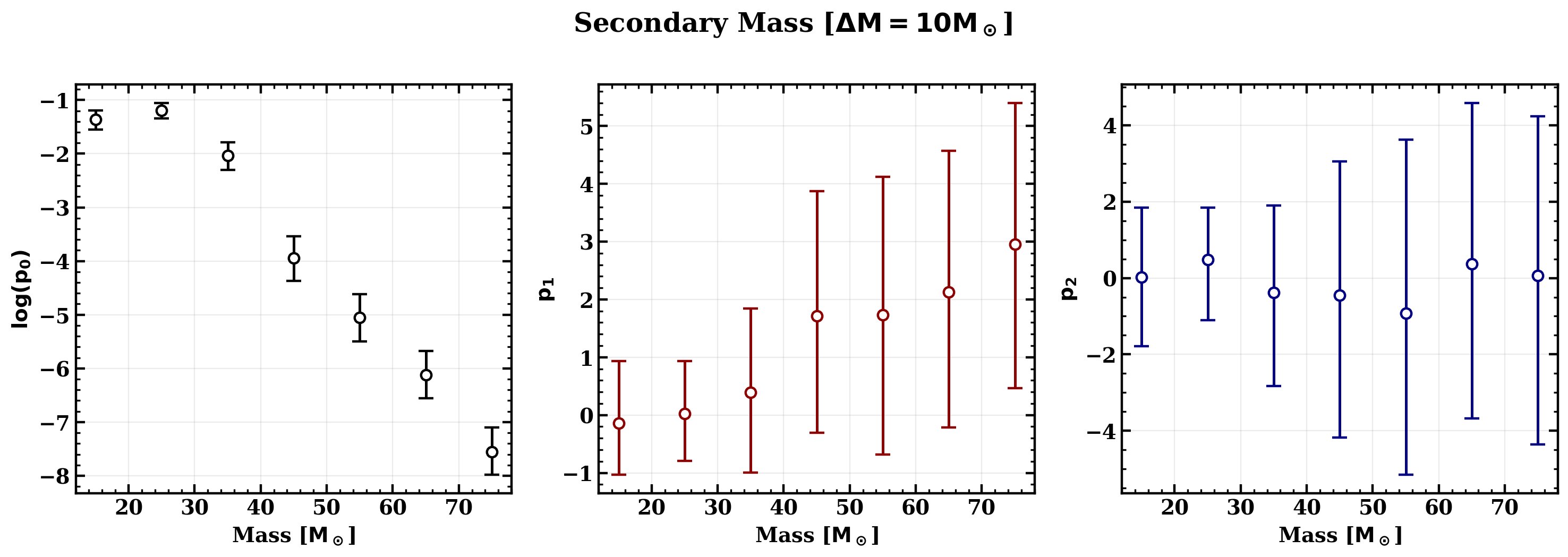}
    \caption{Posterior constraints on the redshift evolution of the BBH secondary mass ($\rm{m_2}$) distribution for a binning scheme with $\rm{\Delta M = 10\,M_{\odot}}$ and $\rm{\Delta z = 0.1}$.  \textbf{Left:} Constant coefficient $\rm{\log(P_0)}$ describing the baseline primary mass distribution. \textbf{Middle:} Linear coefficient $\rm{P_1}$, quantifying the first-order variation of the mass distribution with redshift. \textbf{Right:} Quadratic coefficient $\rm{P_2}$, probing possible curvature in the redshift dependence.  Error bars represent $\rm{1\sigma}$ credible intervals obtained from the hierarchical Bayesian analysis.  The linear coefficient is consistent with zero at low masses ($\rm{m_1 \lesssim 30\,M_{\odot}}$) and shows a mild positive trend toward higher masses, suggesting that more massive black holes may be relatively more abundant at earlier cosmic times. The quadratic coefficient remains consistent with zero across all mass bins, indicating that a linear redshift dependence adequately describes the population within the current observational range $\rm{z \lesssim 1}$ accessible to LVK.}
    \label{fig:Mass1SmallBincoefficients}
\end{figure*}

\begin{figure*}[t]
    \centering
    \includegraphics[width=1.00\textwidth, height=6.0cm]{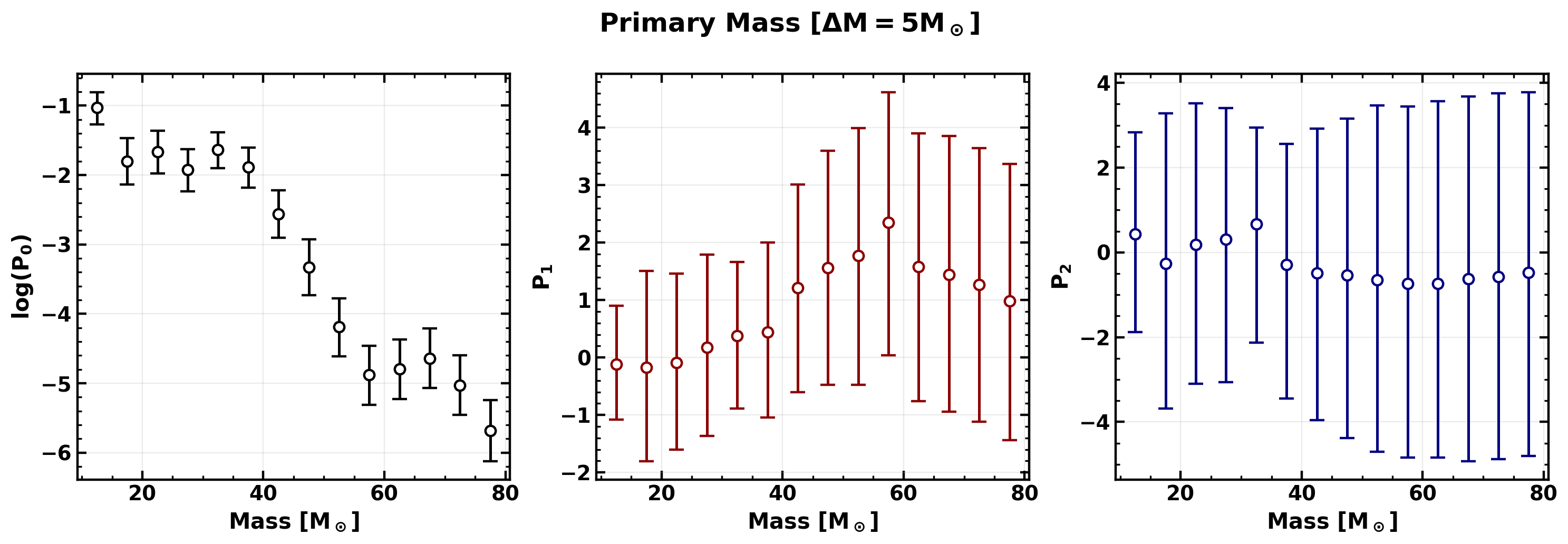}
    \caption{Posterior constraints on the redshift evolution of the BBH primary mass ($\rm{m_1}$) distribution for a binning scheme with $\rm{\Delta M = 5\,M_{\odot}}$ and $\rm{\Delta z = 0.1}$.  \textbf{Left:} Constant coefficient $\rm{\log(P_0)}$ describing the baseline primary mass distribution. \textbf{Middle:} Linear coefficient $\rm{P_1}$, quantifying the first-order variation of the mass distribution with redshift. \textbf{Right:} Quadratic coefficient $\rm{P_2}$, probing possible curvature in the redshift dependence. Error bars represent $\rm{1\sigma}$ credible intervals obtained from the hierarchical Bayesian analysis.  The linear coefficient is consistent with zero at low masses ($\rm{m_1 \lesssim 30\,M_{\odot}}$) and shows a mild positive trend toward higher masses, suggesting that more massive black holes may be relatively more abundant at earlier cosmic times. The quadratic coefficient remains consistent with zero across all mass bins, indicating that a linear redshift dependence adequately describes the population within the current observational range $\rm{z \lesssim 1}$ accessible to LVK.}
    \label{fig:Mass2LargeBincoefficients}
\end{figure*}

\begin{figure*}[t]
    \centering
    \includegraphics[width=1.00\textwidth, height=6.0cm]{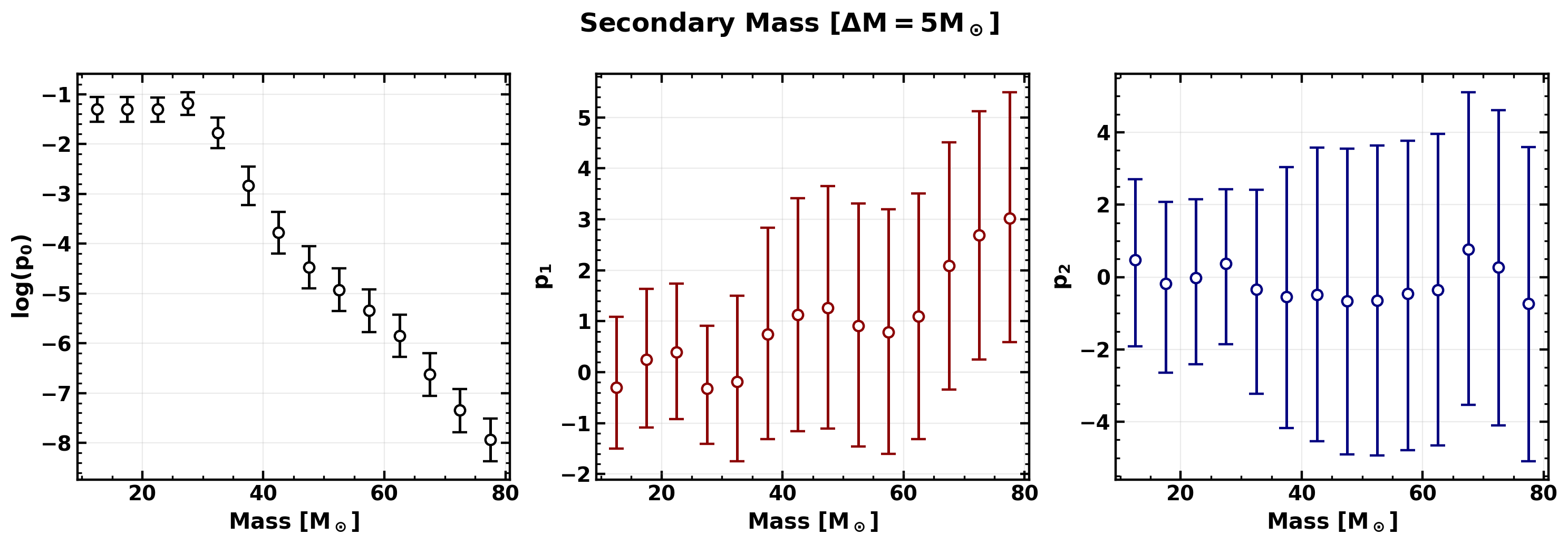}
    \caption{Posterior constraints on the redshift evolution of the BBH secondary mass ($\rm{m_2}$) distribution for a binning scheme with $\rm{\Delta M = 5\,M_{\odot}}$ and $\rm{\Delta z = 0.1}$.  \textbf{Left:} Constant coefficient $\rm{\log(P_0)}$ describing the baseline primary mass distribution. \textbf{Middle:} Linear coefficient $\rm{P_1}$, quantifying the first-order variation of the mass distribution with redshift. \textbf{Right:} Quadratic coefficient $\rm{P_2}$, probing possible curvature in the redshift dependence.  Error bars represent $\rm{1\sigma}$ credible intervals obtained from the hierarchical Bayesian analysis.  The linear coefficient is consistent with zero at low masses ($\rm{m_1 \lesssim 30\,M_{\odot}}$) and shows a mild positive trend toward higher masses, suggesting that more massive black holes may be relatively more abundant at earlier cosmic times. The quadratic coefficient remains consistent with zero across all mass bins, indicating that a linear redshift dependence adequately describes the population within the current observational range $\rm{z \lesssim 1}$ accessible to LVK.}
    \label{fig:Mass2SmallBincoefficients}
\end{figure*}

To distinguish between genuine evolution of the BBH mass distribution and observational biases, we employ the LVK injection campaigns Observing Runs O1-O4. The selection function, together with the full posterior samples of the observed events, is incorporated into our hierarchical Bayesian framework (Section~\ref{sec:Framework}). This framework models the intrinsic distribution $\rm{p(m|z)}$ via Taylor expansion, while explicitly marginalizing over measurement uncertainties and correcting for selection effects. The results of this analysis are presented in Section~\ref{sec:result}.

\section{Results}
\label{sec:result}

We apply the non parametric Bayesian framework described in Section~\ref{sec:Framework} to the GWTC-3 and GWTC-4 catalog of BBH mergers, incorporating the corresponding injection campaigns to account for detector selection effects. The analysis is performed in the $\rm{(m,z)}$ plane, discretized into grids of varying resolutions to study the impact of bin size on the results. We consider two cases: one with $\rm{\Delta m = 10\,M_{\odot}}$ and $\rm{\Delta z = 0.1}$, and another with a finer grid having $\rm{\Delta m = 5\,M_{\odot}}$ and $\rm{\Delta z = 0.1}$. Both cases cover the redshift range $\rm{0 \leq z \leq 1}$ and mass range $\rm{10\,M_{\odot} \leq m \leq 80\,M_{\odot}}$. The detector selection function $\rm{S(m,z)}$ is calibrated from the injection campaign of LVK collaboration.

The Bayesian inference was performed using the \texttt{emcee} \citep{foreman2013emcee} affine-invariant ensemble Markov Chain Monte Carlo (MCMC) sampler. For the finer mass binning configuration ($\rm{\Delta M = 5\,M_\odot}$), we used $200$ walkers and ran the chains for $50\,000$ steps, discarding the first $10\,000$ steps as burn-in. For the coarser binning configuration ($\rm{\Delta M = 10\,M_\odot}$), we used $100$ walkers with $25\,000$ steps and removed the first $5\,000$ steps as burn-in. 
Uniform priors were adopted for all Taylor expansion coefficients describing the mass evolution, with $\rm{\log P_0 \in [-13,0]}$, $\rm{P_1 \in [-20,20]}$, and $\rm{P_2 \in [-30,30]}$.

Figures~\ref{fig:Mass1LargeBincoefficients}-\ref{fig:Mass2SmallBincoefficients} summarize the key results of our non-parametric reconstruction of the redshift evolution of the BBH mass distribution. We present constraints separately for the primary mass ($\rm{m_1}$) and secondary mass ($\rm{m_2}$). For the primary mass ($\rm{m_1}$), a clear qualitative trend emerges in the linear coefficient $\rm{p_1(m_1)}$. In both binning schemes, $\rm{p_1(m_1)}$ is consistent with zero at low masses ($\rm{m_1 \lesssim 25\text{-}30,M_{\odot}}$), indicating no statistically significant redshift evolution in this regime. Toward higher masses, however, $\rm{p_1(m_1)}$ shifts systematically toward positive values. In the large-bin case, this manifests as a smooth rise of the linear coefficient beyond $\rm{\sim 40,M_{\odot}}$, while in the finer binning the trend becomes more structured but remains qualitatively similar. This behavior suggests that more massive primary black holes were relatively more abundant at earlier cosmic times, consistent with astrophysical scenarios in which lower metallicity environments at higher redshift favor the formation of heavier remnants. The full posterior distributions of the inferred parameters, including the corresponding corner plots for both binning configurations, are presented in Appendix~\ref{app:corner}.

In contrast, the quadratic coefficient $\rm{p_2(m_1)}$ remains statistically consistent with zero across nearly all mass bins for both discretizations. Although fluctuations are visible, particularly in the finer binning, they are well within the $\rm{1\sigma}$ uncertainties. This indicates no compelling evidence for strong curvature in the redshift dependence of the primary mass distribution within the currently accessible range $\rm{z \lesssim 1}$ probed by LVK.

A broadly similar picture emerges for the secondary mass ($\rm{m_2}$), albeit with larger statistical uncertainties. For the coarser binning ($\rm{\Delta m = 10,M_{\odot}}$), the linear coefficient $\rm{p_1(m_2)}$ shows a gradual increase toward higher masses, with low-mass bins consistent with zero and high-mass bins favoring positive values. In the finer binning, this increasing trend persists but appears noisier, with larger error bars reflecting the reduced number of events per bin. Nevertheless, the overall pattern is consistent with that observed for the primary mass: a tendency for more massive components to exhibit stronger positive redshift evolution.

The quadratic coefficient for secondary mass $\rm{p_2(m_2)}$ does not display any statistically significant deviation from zero across the mass range. While individual bins fluctuate above or below zero, the uncertainties remain large and encompass the null hypothesis of no quadratic evolution. Therefore, within current observational precision, a linear redshift dependence appears sufficient to describe the secondary mass evolution as well.

Our analysis marginalizes over mass ratio and spin distributions by working directly with component masses and using selection functions from comprehensive injection campaigns. Black hole spins remain poorly constrained in current GW observations, and we do not model their potential redshift evolution. However, the mass-dependent detector sensitivity is dominated by the chirp mass dependence of GW amplitude, with spins contributing subdominant corrections, ensuring our mass evolution inferences are robust to spin distribution uncertainties.

Our findings can be compared with previous population studies using GWTC-3. \citet{Fishbach:2021yvy} found hints of increasing characteristic mass with redshift using parametric models, qualitatively consistent with our mildly positive $\rm{p_1(m)}$ at high masses. \citet{Tiwari:2021yvr} reported results consistent with no evolution across all masses, which aligns with our findings at $\rm{m \lesssim 30\,M_\odot}$ but differs at higher masses where very weak indications of evolution. \citet{Karathanasis:2022rtr} explored metallicity-dependent evolution scenarios and found weak evidence for mass evolution, corroborating our mass-dependent trends. 

\citet{Lalleman:2025xcs} studied the redshift evolution of the BBH mass distribution using GWTC-3, focusing on the Gaussian peak and power-law continuum components of the primary mass distribution. They found no statistically significant evidence for redshift evolution in either component, concluding that a redshift-dependent mass spectrum remains possible but is not required by current data. Our results are broadly consistent with this finding, we find no significant evolution at low masses, and only mild tentative indications at higher masses, while extending the analysis to GWTC-4 with a complementary non-parametric approach.

The inclusion of GWTC-4 data, with its extended redshift reach and increased event count, allows our non-parametric approach to reveal mass-resolved evolutionary features that may have been obscured in earlier parametric analyses or smaller catalogs.

Taken together, the figure demonstrates that the observed BBH population is consistent with a scenario in which the abundance of massive black holes increases modestly with redshift, while lower mass systems remain approximately constant. The mass dependent nature of this evolution strongly points to environmental effects such as metallicity shaping the efficiency of massive black hole formation across cosmic time.

\section{Conclusions}
We have developed and applied a fully non-parametric Bayesian framework to infer the redshift evolution of the BBH mass distribution using all confident BBH detections from the LVK GWTC-3 and GWTC-4 catalogs. By expanding the intrinsic distribution $\rm{p(m|z)}$ as a Taylor series around $\rm{z_{\rm ref}=0.25}$, the method captures evolutionary trends without imposing restrictive parametric assumptions, while rigorously accounting for detector selection effects through injection calibrated efficiencies.

Our analysis yields very mild evidence for mass-dependent evolution. The linear evolution coefficient $\rm{p_1(m)}$ is consistent with zero for lower mass systems ($\rm{m \lesssim 30\,M_\odot}$), but becomes increasingly positive at higher masses. In particular, the highest mass bins show very mild indications that massive black holes may have been relatively more abundant at earlier cosmic times. By contrast, the quadratic term $\rm{p_2(m)}$ remains consistent with zero across all mass ranges, indicating that a linear description suffices up to the redshift range $\rm{z \sim 1}$ currently probed by LVK observations.

These results are qualitatively consistent with theoretical expectations for metallicity driven evolution: low metallicity environments at high redshift suppress stellar winds, allowing the formation of more massive stellar cores and correspondingly heavier black hole remnants. The lack of detectable evolution for lower mass systems suggests that their formation channels are less sensitive to environmental variations, consistent with robust core collapse pathways. The approximately linear trend indicates gradual, metallicity driven changes over cosmic time rather than abrupt evolutionary transitions. This trend is qualitatively consistent with the probability of different formation channels inferred from phase-space analyses, showing that above approximately $45$ M$_\odot$ there is an indication of a different population of BBHs \citep{Afroz:2025ikg}.

Beyond these specific findings, this work demonstrates the power of non-parametric approaches for GW population inference. Unlike parametric models, which may obscure or bias subtle features, our Taylor expansion framework provides a flexible yet interpretable means of capturing population trends while fully correcting for observational selection effects (see Section~\ref{sec:Framework}). The method’s ability to probe possible mass-dependent evolution highlights its value for future analyses with larger datasets.

Previous studies based on GWTC-3 \citep{Lalleman:2025xcs,Fishbach:2021yvy,vanSon:2021zpk,Tiwari:2021yvr,Karathanasis:2022rtr,Rinaldi:2023bbd, Heinzel:2024hva,Lalleman:2025xcs} reported mixed evidence for mass-redshift evolution, ranging from no detectable trends to weak indications at high masses. Some works found hints of increasing characteristic mass with redshift \citep{Fishbach:2021yvy}, while others reported results consistent with no evolution \citep{Tiwari:2021yvr}.

Our combined GWTC-3 and GWTC-4 analysis, employing a complementary non-parametric approach with mass resolved evolution coefficients, finds very mild evidence for evolution at high masses, consistent with the findings of \citet{Lalleman:2025xcs}, while further  corroborating where in mass such evolution may be most pronounced.

Looking forward, the increase in detection rates expected from next generation ground based detectors such as Cosmic Explorer \citep{Reitze:2019iox} and the Einstein Telescope \citep{Branchesi:2023mws} will extend BBH catalogs to higher redshift, enabling reconstruction of higher order evolutionary features with high statistical precision. Our non parametric methodology is well positioned to exploit these datasets, offering both robustness and flexibility in probing the cosmic history of black hole formation. As GW astronomy transitions into the precision era, such model independent frameworks will be essential for extracting the full astrophysical information encoded in BBH populations, ultimately linking GW observations to stellar evolution, galactic chemical enrichment, and the early universe.

\section*{Acknowledgments}
The authors express their gratitude to Souradeep Pal for reviewing the manuscript and providing useful comments as a part of the LIGO publication policy. This work is part of the \texttt{⟨data|theory⟩ Universe Lab}, supported by TIFR and the Department of Atomic Energy, Government of India. The authors express gratitude to the system administrator of the computer cluster of \texttt{⟨data|theory⟩ Universe Lab}. This research is supported by the Prime Minister Early Career Research Award, Anusandhan National Research Foundation, Government of India. We thank the LIGO Scientific Collaboration, the Virgo Collaboration, and the KAGRA Collaboration for providing the gravitational-wave data through the Gravitational Wave Open Science Center (GWOSC). LIGO, funded by the U.S. National Science Foundation (NSF), and Virgo, supported by the French CNRS, Italian INFN, and Dutch Nikhef, along with contributions from Polish and Hungarian institutes. This collaborative effort is backed by the NSF’s LIGO Laboratory, a major facility fully funded by the National Science Foundation. The research leverages data and software from the Gravitational Wave Open Science Center, a service provided by LIGO Laboratory, the LIGO Scientific Collaboration, Virgo Collaboration, and KAGRA. Advanced LIGO's construction and operation receive support from STFC of the UK, Max Planck Society (MPS), and the State of Niedersachsen/Germany, with additional backing from the Australian Research Council. Virgo, affiliated with the European Gravitational Observatory (EGO), secures funding through contributions from various European institutions. Meanwhile, KAGRA's construction and operation are funded by MEXT, JSPS, NRF, MSIT, AS, and MoST. This material is based upon work supported by NSF’s LIGO Laboratory which is a major facility fully funded by the National Science Foundation. We acknowledge the use of the following packages in this work: Numpy \citep{van2011numpy}, Scipy \citep{jones2001scipy}, Matplotlib \citep{hunter2007matplotlib}, Astropy \citep{robitaille2013astropy}, pygtc \citep{Bocquet2016}, and emcee \citep{foreman2013emcee}.

\appendix

\section{Posterior Distributions of Evolution Coefficients}
\label{app:corner}

In this appendix we present the full posterior distributions of the non-parametric evolution coefficients inferred in our hierarchical Bayesian analysis.  The intrinsic mass-redshift distribution is expanded as a Taylor series in redshift,
\begin{equation}
\rm{p(m,z) = p_0(m) + p_1(m)(z-z_{\rm ref}) + p_2(m)(z-z_{\rm ref})^2} ,
\end{equation}

where $\rm{p_0(m)}$ represents the baseline mass distribution, while $\rm{p_1(m)}$ and $\rm{p_2(m)}$ capture the linear and quadratic redshift evolution of the population, respectively. We show the full posterior distributions and parameter correlations using corner plots. These visualizations provide a comprehensive view of parameter uncertainties, degeneracies, and the stability of the hierarchical inference across different binning choices.

\begin{figure*}
\centering
\includegraphics[width=\textwidth]{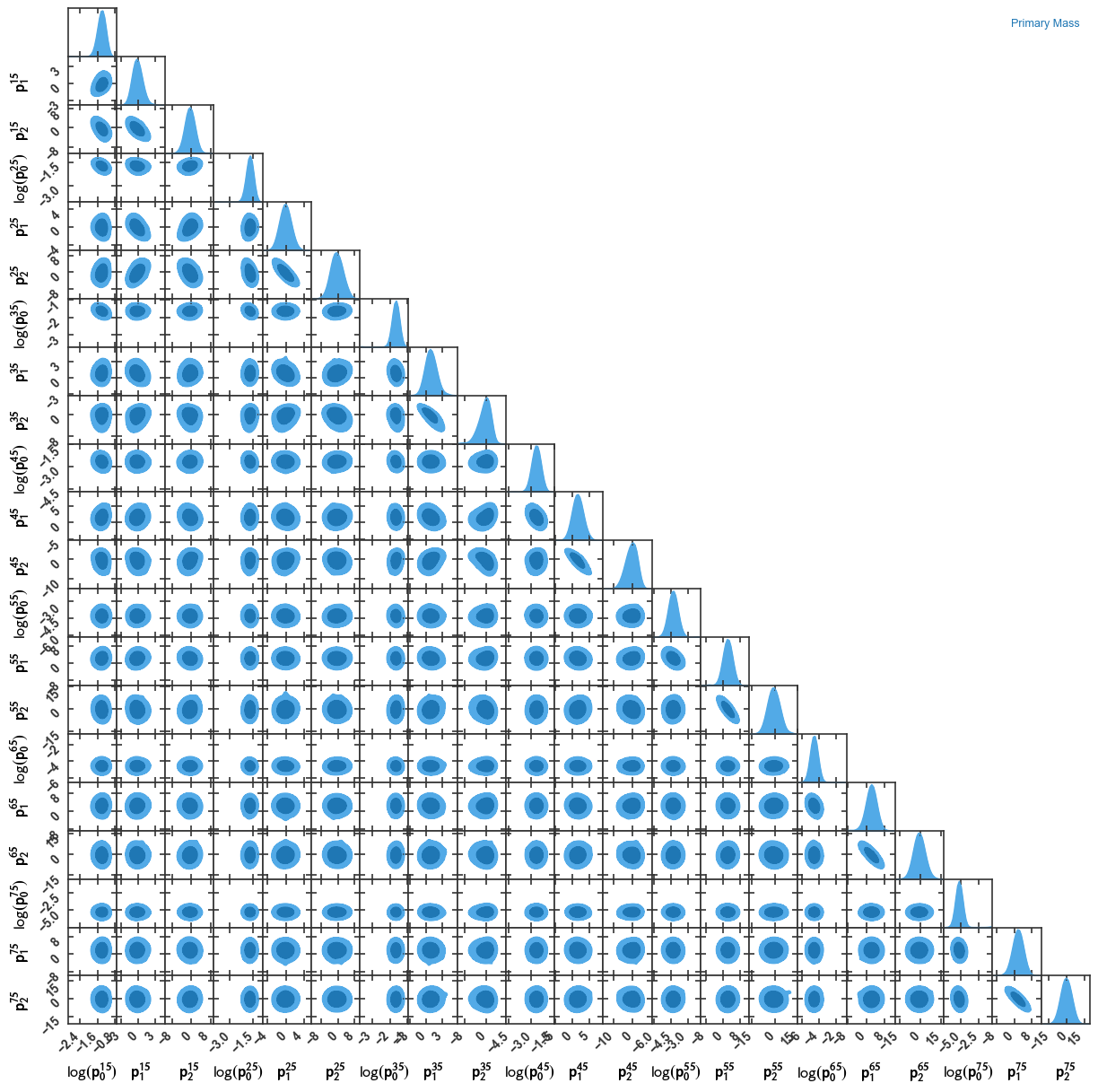}
\caption{Corner plot showing the posterior distributions of the evolution coefficients for the primary black hole mass ($m_1$) using the coarse binning scheme ($\rm{\Delta M= 10 M_\odot}$). The parameters correspond to the Taylor expansion coefficients $\{p_0(m_k), p_1(m_k), p_2(m_k)\}$. The diagonal panels show the marginalized one-dimensional posterior distributions, while the off-diagonal panels show the joint posterior distributions with $1\sigma$ and $2\sigma$ credible regions. This configuration contains 21 parameters corresponding to the coefficients of the non-parametric reconstruction of the mass-redshift distribution.}
\label{fig:corner_primary_large}
\end{figure*}

\begin{figure*}
\centering
\includegraphics[width=\textwidth]{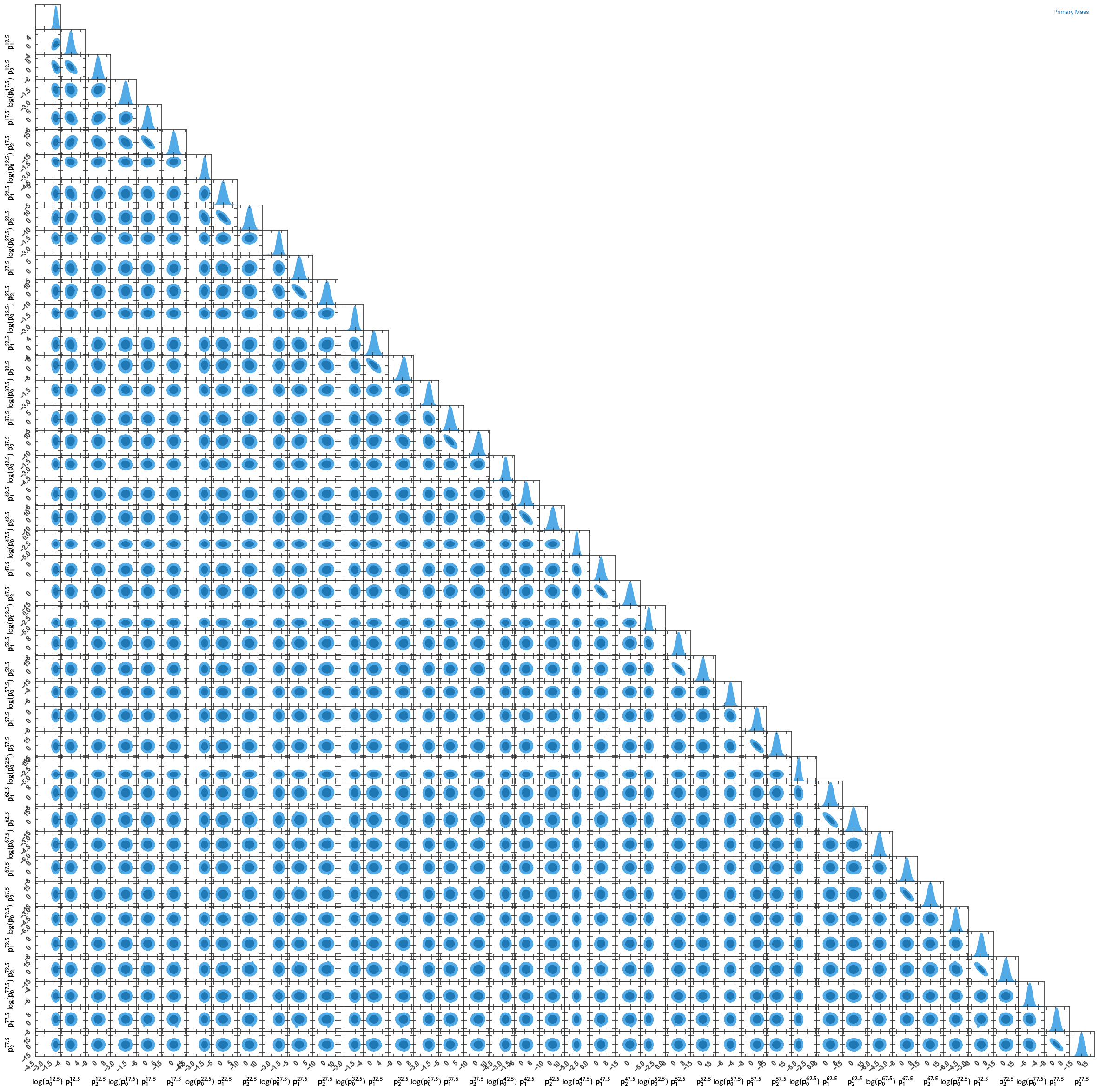}
\caption{Corner plot showing the posterior distributions of the evolution coefficients for the primary black hole mass ($m_1$) using the finer mass binning scheme ($\rm{\Delta M= 5 M_\odot}$). The parameters again correspond to the coefficients $\{p_0(m_k), p_1(m_k), p_2(m_k)\}$. The increased number of bins leads to a higher-dimensional parameter space with 42 parameters in total.}
\label{fig:corner_primary_small}
\end{figure*}

\begin{figure*}
\centering
\includegraphics[width=\textwidth]{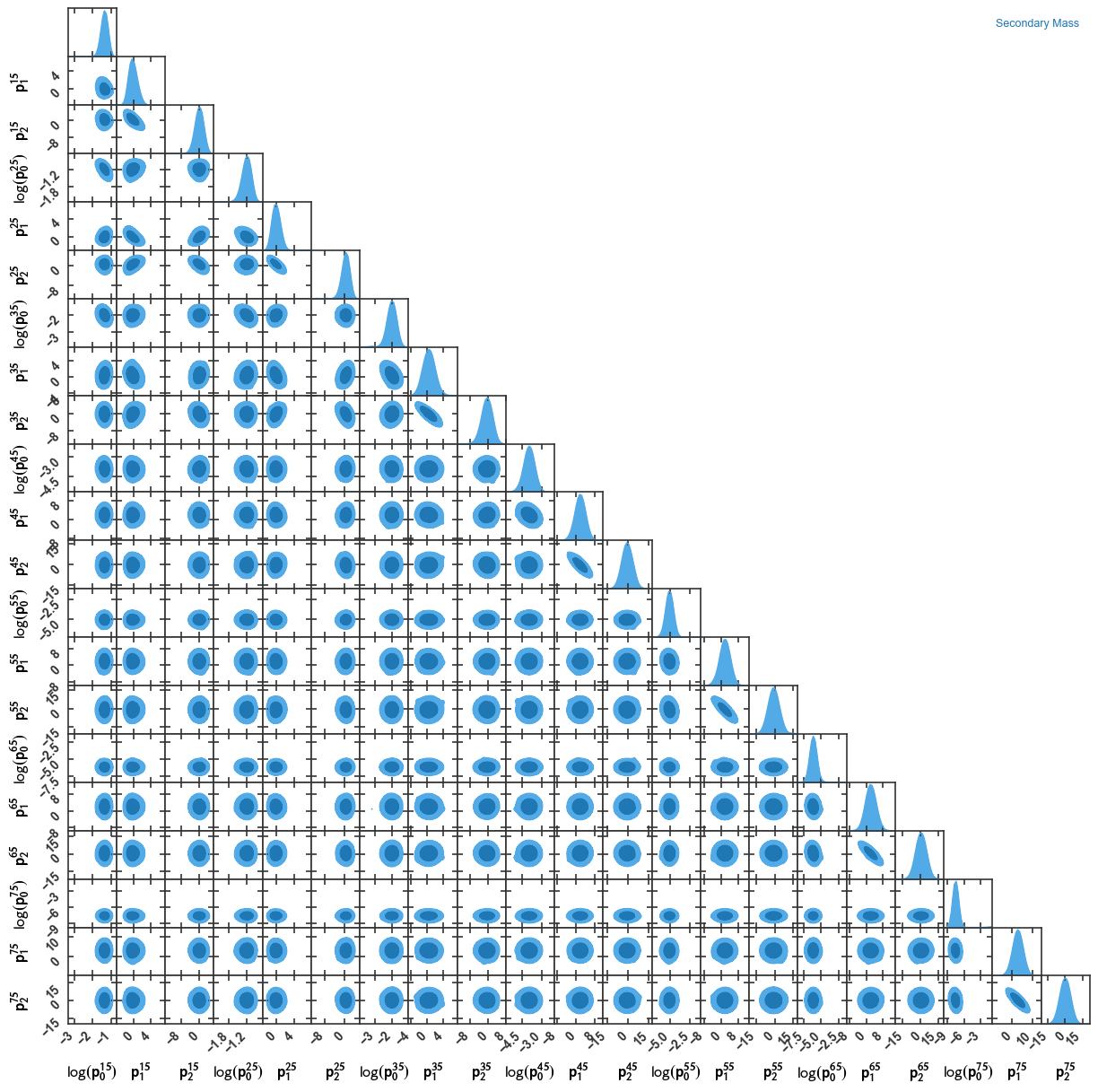}
\caption{Corner plot showing the posterior distributions of the evolution coefficients for the secondary black hole mass ($m_2$) using the coarse binning scheme ($\rm{\Delta M= 10 M_\odot}$). The parameters correspond to the Taylor expansion coefficients $\{p_0(m_k), p_1(m_k), p_2(m_k)\}$ describing the redshift evolution of the secondary mass distribution. As in the primary mass analysis, the diagonal panels show marginalized one-dimensional posteriors while the off-diagonal panels show joint posterior distributions and parameter correlations.}
\label{fig:corner_secondary_large}
\end{figure*}

\begin{figure*}
\centering
\includegraphics[width=\textwidth]{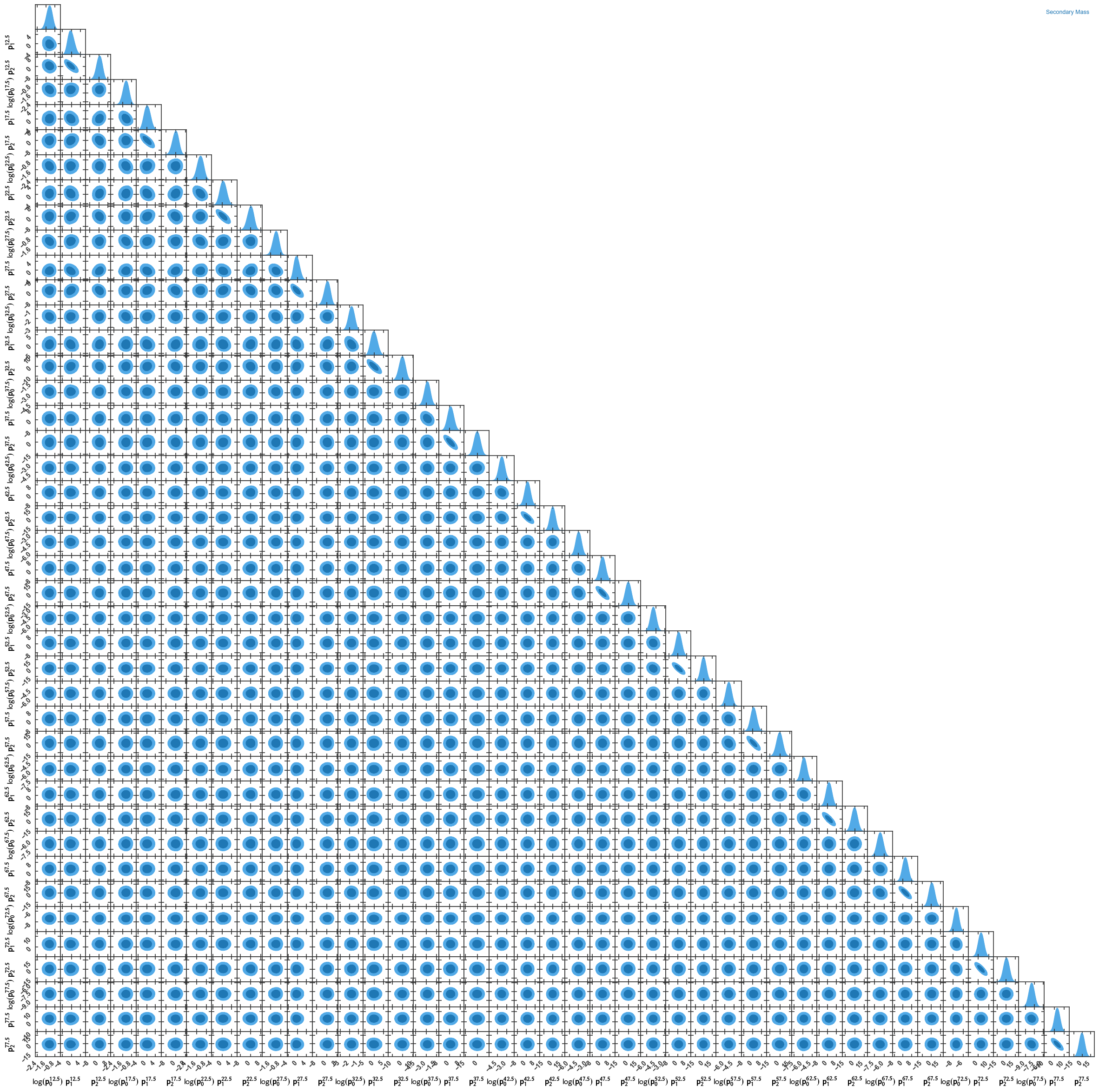}
\caption{Corner plot showing the posterior distributions of the evolution coefficients for the secondary black hole mass ($m_2$) using the finer mass binning scheme ($\rm{\Delta M= 5 M_\odot}$). The larger number of mass bins leads to a higher-dimensional parameter space with 42 parameters. The corner plot illustrates the posterior constraints and correlations among all inferred coefficients in the hierarchical population model.}
\label{fig:corner_secondary_small}
\end{figure*}

Figures~\ref{fig:corner_primary_large} and \ref{fig:corner_primary_small} show the posterior distributions for the primary mass coefficients using coarse ($\rm{\Delta M= 10 M_\odot}$) and fine ($\rm{\Delta M= 5 M_\odot}$) mass binning schemes, respectively. Figures~\ref{fig:corner_secondary_large} and \ref{fig:corner_secondary_small} show the corresponding results for the secondary mass coefficients. The diagonal panels show the marginalized one-dimensional posterior distributions for each parameter, while the off-diagonal panels display the two-dimensional joint posteriors with $1\sigma$ and $2\sigma$ credible regions.

\clearpage

\section{Comparison Between SNR-based and FAR-based Selection Functions}
\label{app:far_vs_snr}

In the main analysis the selection function $\rm{S(m,z)}$ is constructed by treating an injection as detected whenever $\rm{SNR > 10}$. Since the effective SNR threshold corresponding to a fixed FAR varies with source mass, and our analysis explicitly targets mass-dependent features of the population, we verify here that this approximation does not bias the inferred coefficients by repeating the full inference with a FAR-based selection function. An injection is classified as detected if the \texttt{PyCBC} search pipeline reports $\rm{FAR < 1\,yr^{-1}}$, matching the threshold used to define the confident BBH samples in GWTC-3 and GWTC-4 \citep{KAGRA:2021vkt,LIGOScientific:2025slb}. All other elements of the pipeline are held fixed.

\begin{figure*}[t]
\centering
\includegraphics[width=\textwidth]{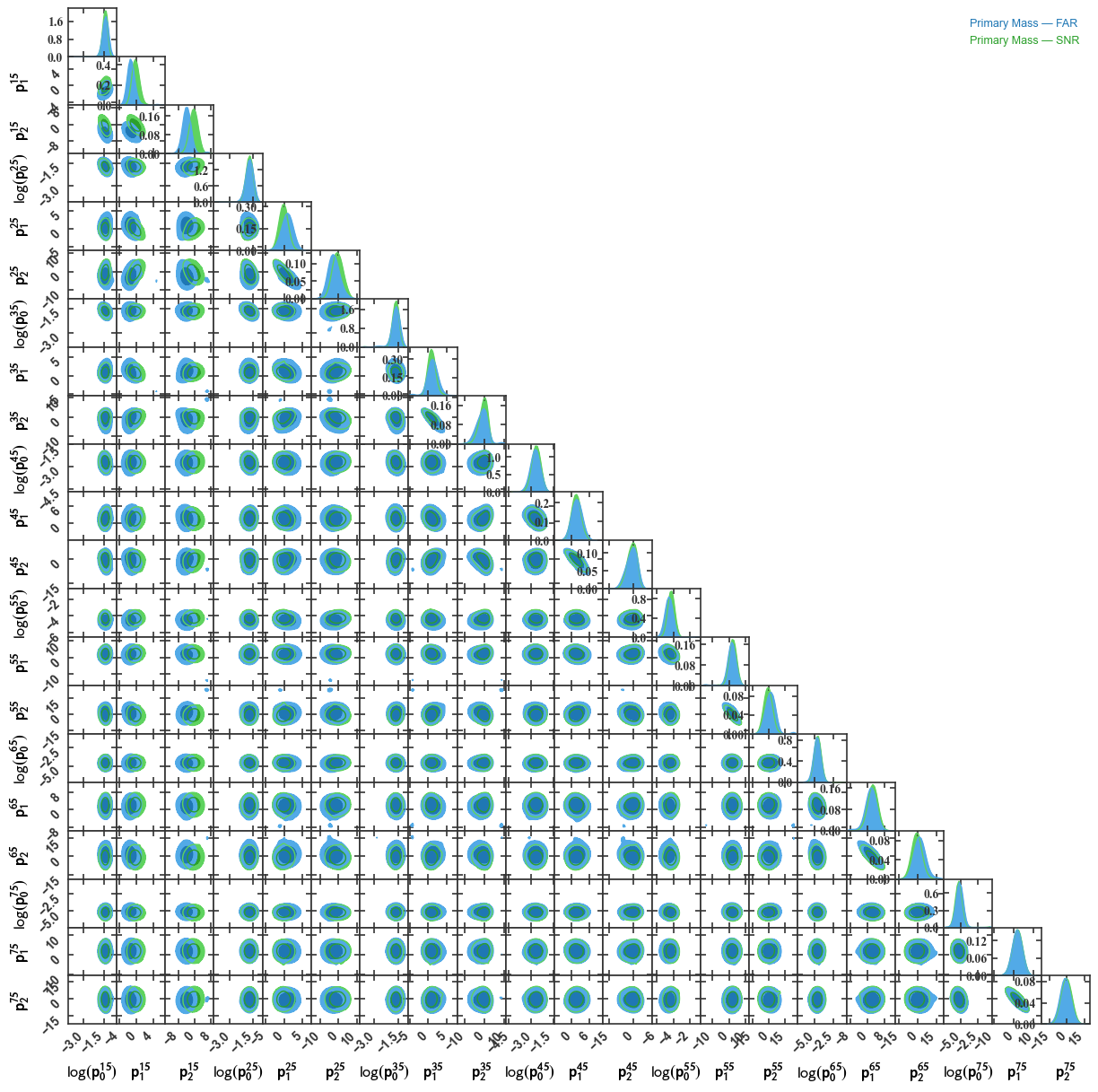}
\caption{Posterior distributions of the evolution coefficients $\rm{\{\log p_0(m_k), p_1(m_k), p_2(m_k)\}}$ for the primary mass $\rm{m_1}$ under the SNR-based ($\rm{SNR > 10}$, green) and FAR-based ($\rm{FAR < 1\,yr^{-1}}$, blue) selection functions, for the $\rm{\Delta M = 10\,M_\odot}$ binning.}
\label{fig:far_vs_snr_primary}
\end{figure*}

\begin{figure*}[t]
\centering
\includegraphics[width=\textwidth]{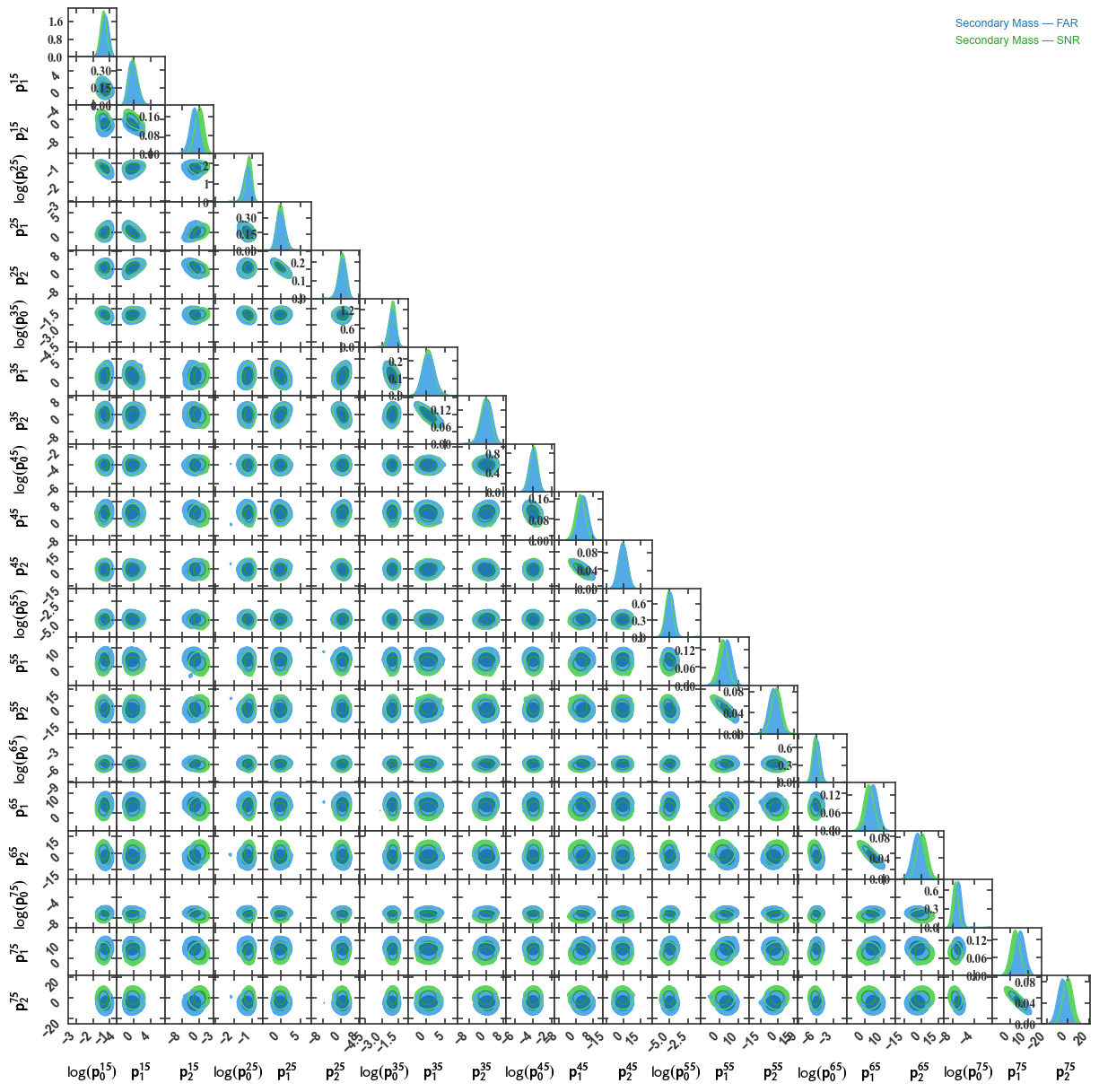}
\caption{Posterior distributions of the evolution coefficients $\rm{\{\log p_0(m_k), p_1(m_k), p_2(m_k)\}}$ for the secondary mass $\rm{m_2}$ under the SNR-based ($\rm{SNR > 10}$, green) and FAR-based ($\rm{FAR < 1\,yr^{-1}}$, blue) selection functions, for the $\rm{\Delta M = 10\,M_\odot}$ binning.}
\label{fig:far_vs_snr_secondary}
\end{figure*}

Figures~\ref{fig:far_vs_snr_primary} and \ref{fig:far_vs_snr_secondary} compare the posterior distributions of the Taylor coefficients $\rm{\{\log p_0(m_k), p_1(m_k), p_2(m_k)\}}$ under the SNR-based (green) and FAR-based (blue) selection functions for the primary and secondary masses, using the $\rm{\Delta M = 10\,M_\odot}$ binning. The two reconstructions are statistically consistent across all 21 parameters of each component. We conclude that the $\rm{SNR > 10}$ threshold provides a good approximation to the FAR-based selection at the current catalog size and does not affect our conclusions, though a fully FAR-based treatment will become necessary as statistical uncertainties shrink in future runs.

\section{ROBUSTNESS TO THE INJECTED SPIN DISTRIBUTION}
\label{app:spin_reweight}

The selection function $S(m,z)$ used throughout this work is estimated by binning recovered injections in the $(m_1,z)$ and $(m_2,z)$ planes, with all remaining parameters marginalized out. Among these marginalized parameters are the component spins. The mass-averaged selection function therefore inherits whatever spin distribution the injection campaign was drawn from. Since the injections adopt broad spin distributions chosen for efficient Monte Carlo coverage rather than for astrophysical plausibility, it is important to verify that our inferred evolution coefficients are not sensitive to this choice. In this appendix we test this explicitly by reweighting the injected spins to a fixed, physically motivated population and recomputing $S(m,z)$. Throughout this test we use the FAR-based selection function ($\mathrm{FAR} < 1\,\mathrm{yr}^{-1}$), and the baseline against which we compare is the same FAR-based analysis presented in Appendix~\ref{app:far_vs_snr}; the reweighting described below is applied on top of that analysis, so that the two differ only in the assumed spin distribution.

The reweighting proceeds injection by injection. Each injection is given a weight equal to the probability of its spin parameters (both magnitudes and both tilts) under a plausible astrophysical population, divided by the probability under the distribution from which it was drawn. For the target population we adopt the default component spin model of the GWTC-4.0 population analysis, with spin magnitudes following a truncated Gaussian and tilts a mixture of an isotropic and a preferentially aligned component. This removes the injected spin distribution and imposes the plausible one in its place. With unit weights the procedure recovers the FAR-based selection function of the main analysis, so baseline and reweighted cases differ only in the assumed spin distribution; all other elements of the pipeline are held fixed.

\begin{figure*}
\centering
\includegraphics[width=\textwidth]{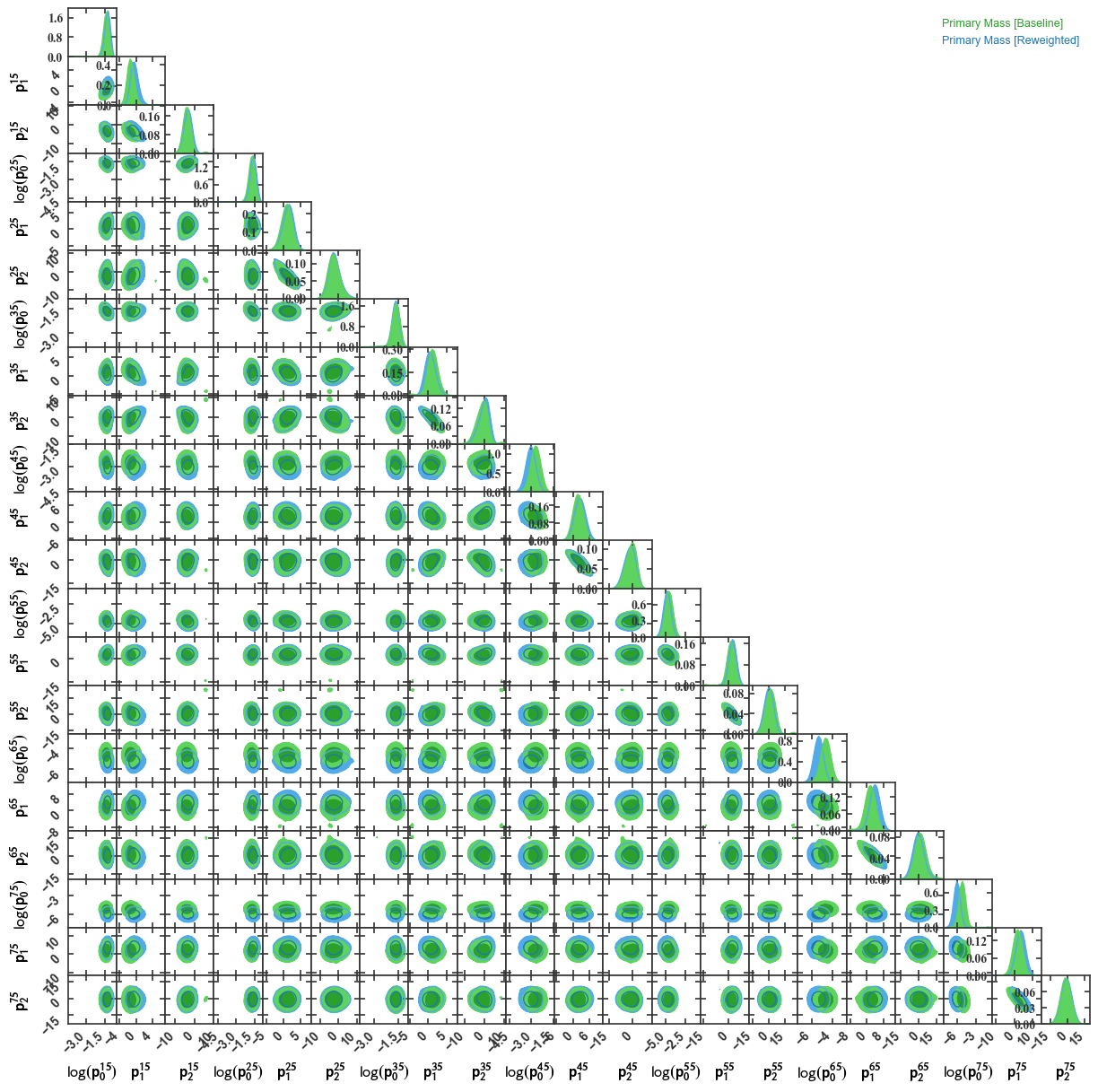}
\caption{Joint posterior distributions of the evolution coefficients $\{\log p_0(m_k), p_1(m_k), p_2(m_k)\}$ for the primary mass $m_1$, comparing the baseline FAR-based analysis with the injected spin distribution (green) against the analysis with spins reweighted to a plausible astrophysical population (blue), for the $\Delta m = 10\,M_\odot$ binning. The green distribution is identical to the FAR-based reconstruction shown in Appendix~\ref{app:far_vs_snr}. The two reconstructions are consistent across all parameters, demonstrating that our results are insensitive to the injected spin distribution.}
\label{fig:spin_reweight_m1}
\end{figure*}

\begin{figure*}
\centering
\includegraphics[width=\textwidth]{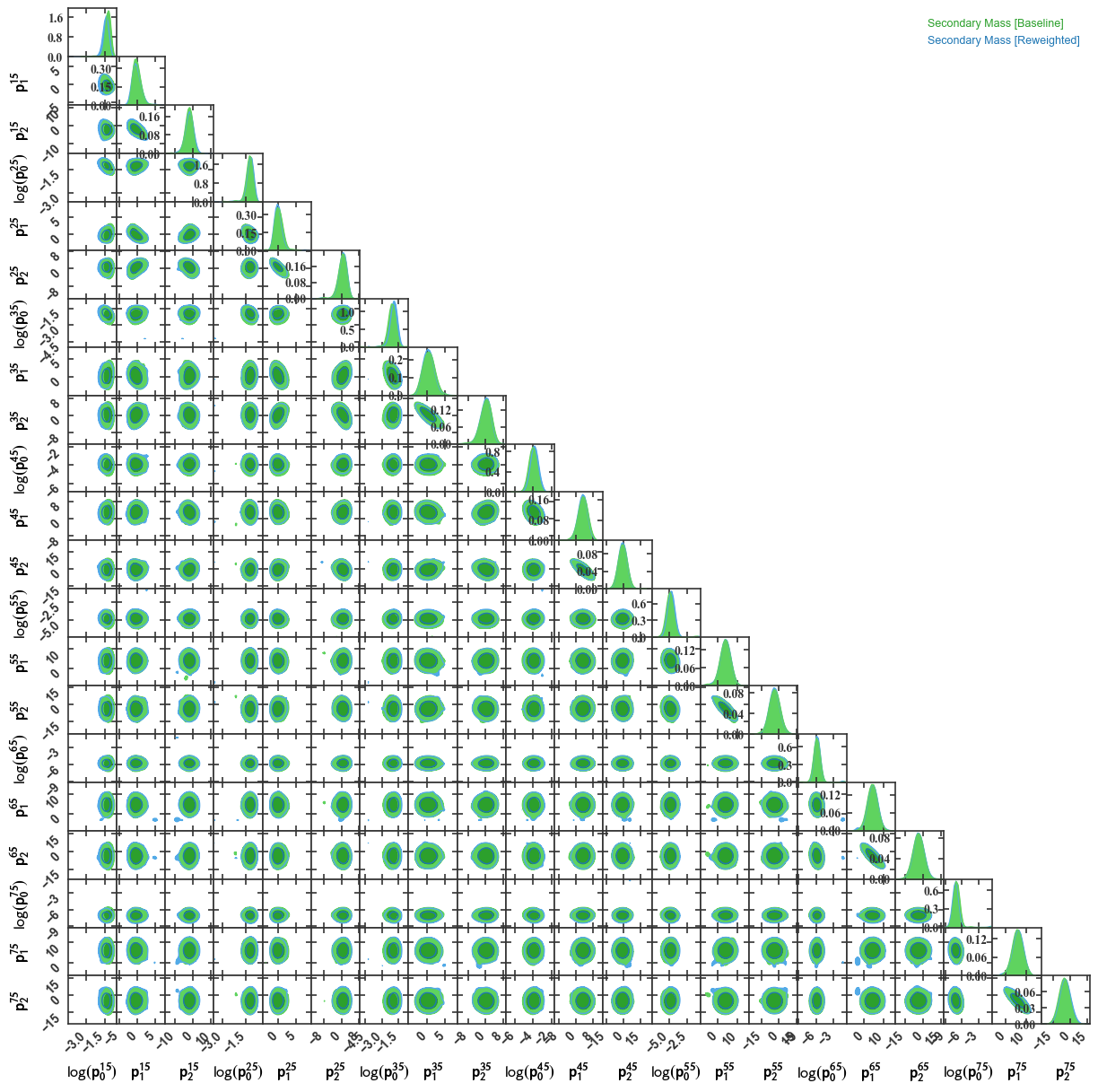}
\caption{As Figure~\ref{fig:spin_reweight_m1}, but for the secondary mass $m_2$. The baseline and reweighted posteriors agree across all parameters.}
\label{fig:spin_reweight_m2}
\end{figure*}

Figures~\ref{fig:spin_reweight_m1} and~\ref{fig:spin_reweight_m2} show the full joint posterior distributions of the Taylor coefficients $\{\log p_0(m_k), p_1(m_k), p_2(m_k)\}$ obtained with the baseline (injected) spins and with the reweighted spins, for the primary and secondary masses respectively, using the $\Delta m = 10\,M_\odot$ binning. The baseline
(green) is identical to the FAR-based reconstruction of Appendix~\ref{app:far_vs_snr}. The two reconstructions are statistically consistent across all parameters for both components: the marginalized posteriors overlap well within their $1\sigma$ credible regions.

We conclude that the mass-dependent evolution reported in this work is robust to the assumed spin distribution of the injection campaigns, consistent with our expectation that the mass dependence of the detector sensitivity is dominated by the chirp-mass dependence of the gravitational-wave amplitude, with spins contributing only subdominant corrections.

\clearpage
\section*{Data Availability}
\label{sec:dataavail}
The gravitational wave catalogs used in this study are publicly available 
on Zenodo as part of the LVK data releases: \citet{GWTC2p1_zenodo}, 
\citet{GWTC3_zenodo}, and \citet{GWTC4_zenodo}.

\bibliography{references}
\end{document}